\documentclass{article}
\usepackage{epsfig}
\begin{document}
\vspace*{2cm}
\begin{center}
{ \Large \bf RANDOM PHASE APPROXIMATION \\ \vskip 0.2 cm
AND NEUTRINO-NUCLEUS CROSS SECTIONS  
\footnote{Presented at the {\sl XX Max Born Symposium}, 
 Dec. 7-10 2005, Wroc{\l}aw (Poland)}
}

\vspace{1.5cm}
{\large Giampaolo Co'} \\
{Dipartimento di Fisica,  Universit\`a di Lecce, \\
I-73100 Lecce, Italy} \\
\vspace{.5cm}
{Istituto Nazionale di Fisica Nucleare  sez. di Lecce,
\\ I-73100 Lecce, Italy} \\

\vspace{.5cm}

\end{center}

\begin{abstract}
  The Random Phase Approximation theory is used to calculate the total
  cross sections of electron neutrinos on $^{12}$C nucleus.  The role
  of the excitation of the discrete spectrum is discussed.  A
  comparison with electron scattering and muon capture data is
  presented. The cross section of electron neutrinos coming from muon
  decay at rest is calculated.
\end{abstract}


\vskip 0.5 cm 

The Random Phase Approximation (RPA) is an effective theory
constructed to study the excitations of many-body systems.  The RPA
assumes that the exited states of these systems can be described as
linear combinations of one-particle on-hole ($1p-1h$) and one-hole
one-particle ($1h-1p$) excitations. The goal of the theory is to find
the coefficients of the linear combinations for a given interaction
between particles and holes.

In nuclear physics, the RPA has been applied to study excitations on a
wide energy range, from a few MeV, the discrete spectrum, up to
hundreds of MeV, in a regime called quasi-elastic where the emission
of a single nucleon is the dominant process. One of the great
successes of the RPA is the prediction of collective surface
vibrations, called giant resonances, appearing at energies between 15
and 30 MeV in all the nuclei with more than 10 nucleons.

\begin{figure}[ht]
\begin{center}
\includegraphics [angle=90, scale=0.45]
                 {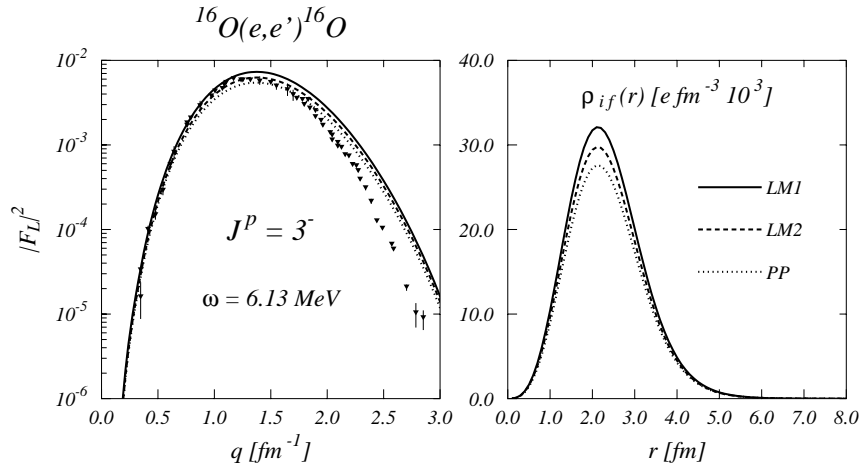}
\caption{\small 
Charge form factors (left panel) and transition densities (right
panel) of the low lying 3$^-$ state in $^{16}$O at 6.13 MeV calculated
in RPA by using three different interactions which
reproduce the excitation energy value. The data are form Ref. 
\protect\cite{but86}.
}
\label{fig:ff3m} 
\end{center}
\end{figure}

The inputs required by the RPA are the set of single particle energies
and wave functions, and the effective interaction between particles
and holes.  In our calculations the single particle basis, which
properly includes the continuum, has been obtained by solving the
one-body Schr\"odinger equation with a spherical Woods-Saxon
potential.  The parameters have been taken from the literature
\cite{bot05a}, and have been fixed to reproduce the rms charge radii
and the single particle energies close to the Fermi level.  The
theoretical uncertainty has been studied by using various $ph$
interactions, specifically the LM1, LM2, and PP interactions of Ref.
\cite{bot05a}. The LM1 and LM2 interactions are zero-range forces of
Landau-Migdal type with slightly different values of the parameters.
The PP interaction is a finite-range interaction. A common
characteristic of the three interactions is that they have been
rescaled to reproduce the excitation energy of the low lying 3$^-$
state in $^{16}$O at 6.13 MeV. Even though the three interactions
produce the same excitation energy, they give different descriptions
of the 3$^-$ state. This is shown in Fig. \ref{fig:ff3m} where we
compare the 3$^-$ charge form factors with the data of Ref.
\cite{but86} measured in inelastic electron scattering experiments.
The difference between the various results indicates the magnitude of
the theoretical uncertainty.
%
%

In recent years, we have used the continuum RPA to study neutrino
induced excitations of the $^{12}$C and $^{16}$O nuclei in the
discrete low-lying region, in the giant resonance and in the
quasi-elastic regions.  Since a detailed presentation of our results
in the giant resonance and in the quasi-elastic regions can be found
elsewhere \cite{bot05a, bot05b, ble01, co02, co06}, in this
contribution we discuss the role of the discrete excitation region.
%
%
\begin{table} [ht]
\vskip 0.5 cm 
\begin{center}
\begin{tabular}{lccc}
\hline
     & $^{12}C$ & $^{12}N$ & $^{12}B$ \\
\hline
LM1  & 17.2 & 20.2 & 14.3 \\
LM2  & 18.8 & 21.7 & 15.9 \\
PP   & 16.7 & 19.6 & 13.8 \\
NI05 & 15.1 & 18.0 & 12.2 \\
exp  & 15.1 & 17.3 & 13.4 \\
\hline
\end{tabular}
\end{center}
\caption 
{\small Energies, in MeV, of the isospin triplet 1$^+$ excited
states referred to the $^{12}$C ground state.
}
\label{tab:onep}
\end{table}
%
%
%
\begin{figure}[ht]
\begin{center}
\includegraphics [angle=90, scale=0.45]
                 {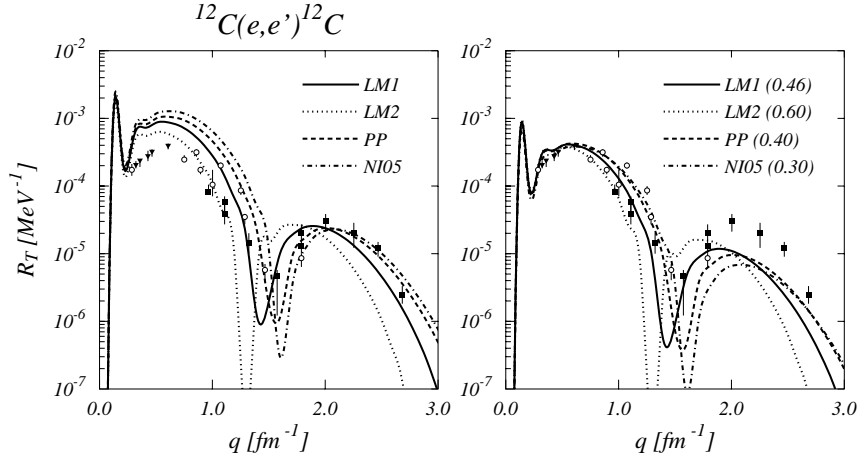}
\caption{\small 
Magnetic form factors of the  1$^+$  in $^{12}$C calculated with 
various interactions and compared with the data of Ref.
\protect\cite{don68}. The results of the left panel has been
obtained by using the quenching factors indicated in the labels.
}
\label{fig:ff1p} 
\end{center}
\end{figure}

In the evaluation of the total neutrino cross section, the excitation
of bound states cannot be neglected, even for neutrino energies well
above the continuum emission threshold. As an example of the relevance
of the discrete excitations, we consider here the case of the isovector
triplet formed by the 1$^+$ excited states in the spectrum of
$^{12}$C.
 
The energies of both charge conserving and charge exchange excitations
obtained in RPA are compared in Tab. \ref{tab:onep} with the
experimental values. None of the interactions above presented, well
tuned to describe natural parity states, is able to reproduce the
experimental energies.  We have rescaled the LM1 interaction to obtain
the energy of the charge conserving 1$^+$ state.  Even with this
interaction, that we called NI05, the RPA cannot reproduce the
energies of the charge exchange states.

The difficulties of the RPA in the description of unnatural parity
excitations are even better shown in Fig. \ref{fig:ff1p} where we
compare the magnetic form factors of the 1$^+$ state at 15.1 MeV with
the data of Ref. \cite{don68} measured in inelastic electron
scattering experiments.  The bare RPA results are shown in the left
panel. The spreading of these results is much larger than that shown
in Fig. \ref{fig:ff3m} for the 3$^-$ state of $^{16}$O. In any case,
none of the curves reproduces the peak of the experimental form
factor, where the data are more reliable.  This is a common feature of
almost all the electron scattering magnetic form factors in
medium-heavy nuclei \cite{co90,mok00}.  The source of this deficiency
of the RPA has been widely investigated and various studies indicate
that the problem has to be ascribed to the restriction of the
configuration space to $1p-1h$ excitations \cite{kre80} rather than to
the absence of correlations of short-range type \cite{mok00}.

We overcome this difficulty of the RPA by using quenching factors
whose values are fixed to best fit the peak of the data. In the right
panel of Fig. \ref{fig:ff1p} we show the curves obtained in this
manner. The values of the quenching factors are given in the labels.
%
%
\begin{figure}[ht]
\begin{center}
\includegraphics [angle=0, scale=0.45]
                 {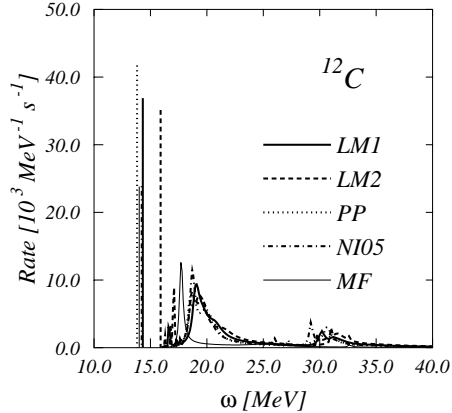}
\caption{\small 
Muon capture rates calculated with various interactions 
as a function of the nuclear excitation energy.
}
\label{fig:muon} 
\end{center}
\end{figure}
%
%
%
\begin{table} [ht]
\begin{center}
\begin{tabular}{lrrrrrr}
\hline
       & LM1   & LM2   & PP   & NI05 & MF    & exp \\ 
\hline
 1$^+$    & 34.93 & 33.86 & 34.56 & 31.57& 23.83 & 6.04 \\
 1$^+$(q) & 16.07 & 20.31 & 13.82 & 9.47 &       &      \\
 2$^+$ &  0.26 &  0.20 &  0.35 &  0.39  &  0.46 & 0.21 \\
 2$^-$ &  1.40 &  0.79 &  0.52 &  8.09  &  9.04 & 0.18 \\
 1$^-$ &  0.32 &  0.29 &  6.31 &  0.98  &  1.13 & 0.62 \\
\hline
dis    & 36.90 & 35.13 & 41.73 & 41.03  & 34.46 & 7.05  \\
dis(q) & 18.04 & 21.48 & 20.99 & 18.93  &       &       \\
con    & 31.35 & 37.09 & 28.28 & 31.16  & 12.48 & 30.04 \\
tot    & 68.25 & 72.22 & 70.53 & 72.19  & 46.94 & 37.09 \\
tot(q) & 50.39 & 58.67 & 49.79 & 50.09  &       &   \\
\hline
\end{tabular}
\end{center}
\vskip 1.0 cm
\caption{\label{tab:muon} 
\small Muon capture rates of $^{12}$C in 10$^3$ s$^{-1}$. 
The rows labeled with (q) shows the results obtained by using the
quenching factors given in Fig. \protect\ref{fig:ff1p}.
The experimental values have been taken from Ref.  
\protect\cite{mea01}.
}
\end{table}

The muon capture is an interesting case where the excitation
of both bound and continuum states should be considered.  In Fig.
\ref{fig:muon} we show the muon capture rate of the $^{12}$C nucleus
calculated with the various interactions, as a function of the nuclear
excitation energy.  The largest contribution to the total rate is
given by the excitation of the continuum, dominated by the giant
resonances, but the contribution of the bound states cannot be
neglected.  In the figure the result obtained by a mean-field (MF)
calculation is represented by the full thin line. While all the other
results predict a collective resonance behavior at about 19 MeV, the
MF results does not show this characteristic.

Experimentally, it has been possible to disentangle the contribution
of various $^{12}$C excited states to the total capture rate.  In Tab.
\ref{tab:muon}.  we compare our capture rates with the data quoted in
Ref. \cite{mea01}. Our calculations describe reasonably well the
contribution of the continuum but they overestimate the capture rates
of the discrete states.  The main source of error is due to the 1$^+$
excitation.  For this reason we show in the rows labeled with (q) the
results obtained by multiplying the 1$^+$ states with the quenching
factors of Fig. \ref{fig:ff1p}. The use of the quenching factors
slightly improves the agreement with the data.

By looking to the total rates only, the mean field calculation
provides the value that is in better agreement with the experimental
one.  However, a more detailed analysis of the various contributions
shows that this result is obtained as a sum of a too large discrete
contribution with the too small contribution of the continuum.
%
%
\begin{figure}[ht]
\begin{center}
\includegraphics [angle=90, scale=0.45]
                 {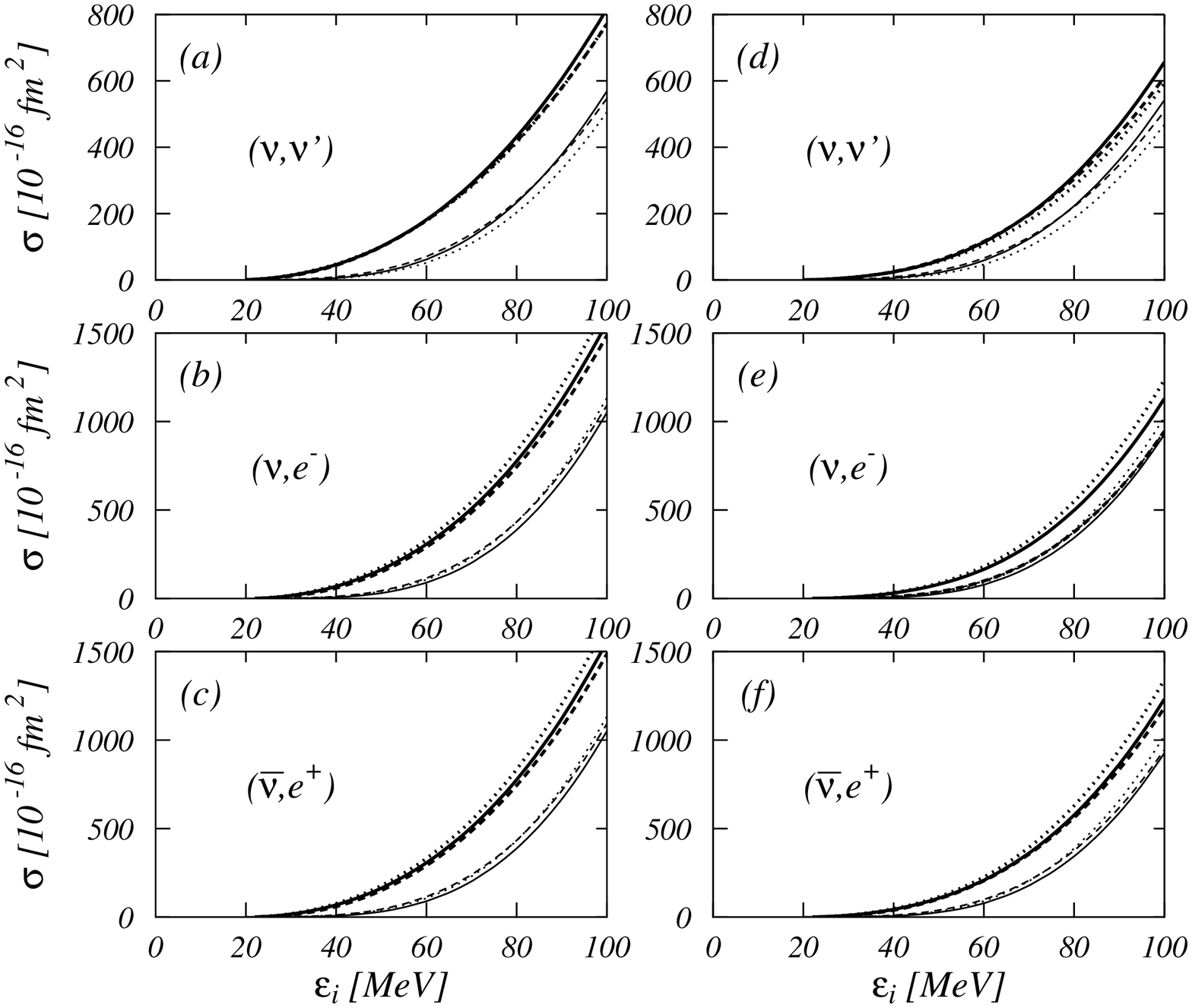}
\caption{\small 
  Total neutrino cross sections as a function of the neutrino
  energies, calculated with various interactions. The full lines show
  the results obtained with the LM1, the dashed lines with the LM2,
  the dotted lines with the PP interaction respectively. The lower,
  thinner, lines have been obtained by considering only the excitation
  to the continuum. The upper, thicker lines include also the
  contribution of the discrete excitation. The left panels show the
  bare RPA results, the right panels the results obtained by using the
  quenching factors of Fig. \protect\ref{fig:ff1p} for the discrete
  excitations.
}
\label{fig:totc12} 
\end{center}
\end{figure}
%
%
%
%
\begin{figure}[ht]
\begin{center}
\includegraphics [angle=90, scale=0.45]
                 {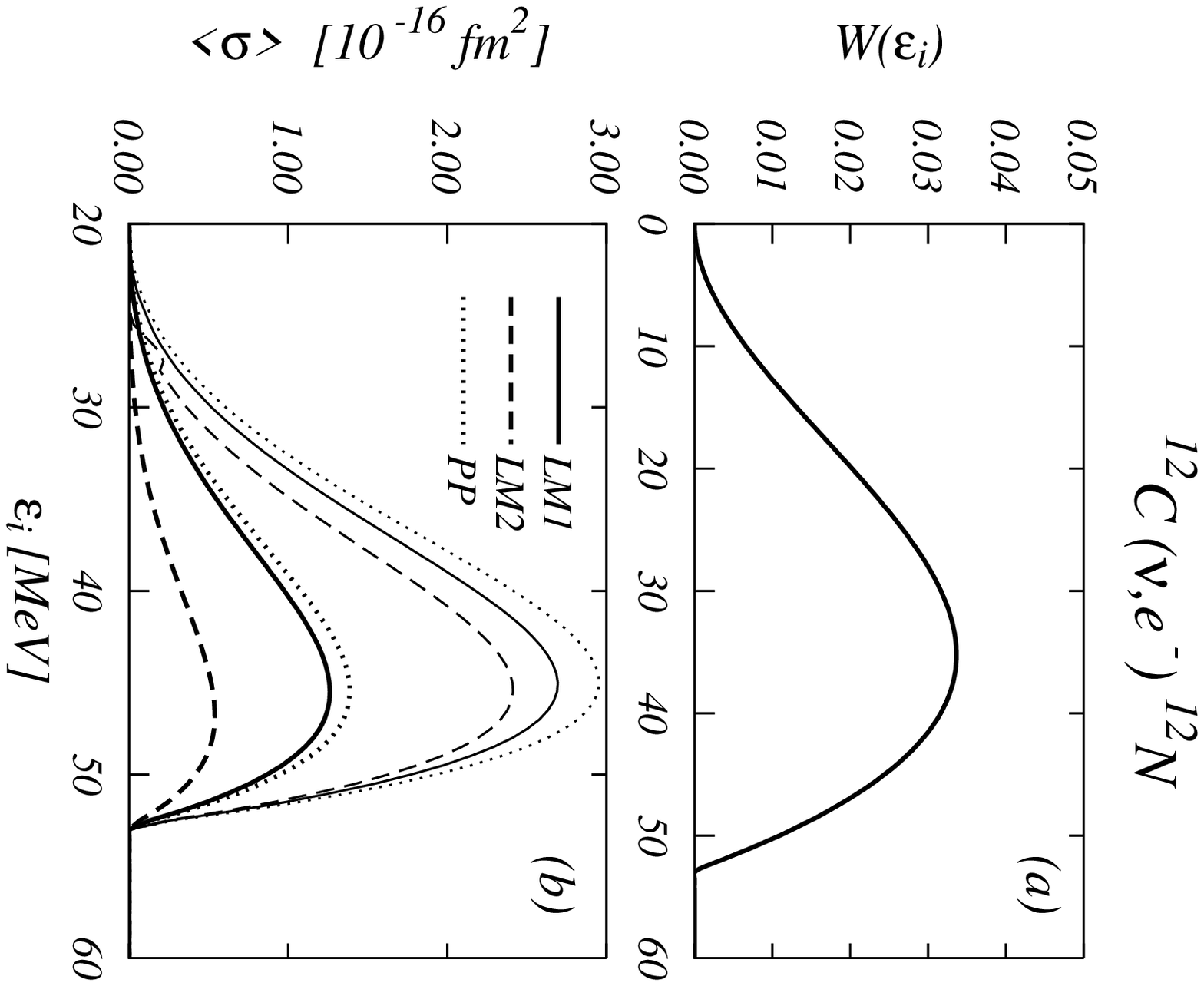}
\caption{\small 
The upper panel shows the energy distribution of the electron
neutrinos coming from muon decay at rest, Michel spectrum. In the
lower panel we show the neutrino the cross sections obtained by
multiplying the energy distribution $W(\epsilon_i)$ with the total
cross sections of the panels (b) and (e) of Fig. 
\protect\ref{fig:totc12}. The upper, thinner, lines show the RPA
results, the lower, thicker, lines have been obtained 
by using the quenching factors  of 
Fig. \protect\ref{fig:ff1p}. 
}
\label{fig:michel} 
\end{center}
\end{figure}

In Fig. \ref{fig:totc12} we show the total neutrino cross sections for
charge conserving and charge exchange reactions as a function of the
neutrino energy.  The lower lines have been obtained by
considering only the excitation to the continuum, while the upper 
curves include also the excitation of the discrete states,
whose main contribution is given by the 1$^+$ state. The lines of the
left panel are those obtained with the bare RPA calculations, while in
those of the right panels the discrete contributions have been
multiplied by the quenching factors of Fig. \ref{fig:ff1p}.
Even in this case the contribution of the discrete excitation is not
negligible. The spreading of the various results is a measure of the
theoretical uncertainty.

As an example of the consequences of this uncertainty, we have
calculated the 
$^{12}$C$(\nu_e,e^-)^{12}$B total sections for 
neutrinos emitted by $\mu^+$ at rest, since this quantity has been
measured \cite{bod94}.
We consider the electron neutrinos coming form the decay
\[
\mu^+ \rightarrow {\overline \nu}_\mu\,\,\, + \nu_e +\,\,\, e^+
\,\,.
\]
The $\nu_e$ are emitted with the energy distribution shown in the
upper panel of Fig. \ref{fig:michel}. The cross sections as a function
of the neutrino energy are calculated by multiplying $W(\epsilon_i)$
with the cross sections of panels (b) and (e) of Fig.
\ref{fig:totc12}. These cross sections are shown in the lower panel of
Fig. \ref{fig:michel}, where the upper, thinner, lines indicate the
bare RPA cross sections, and the lower, thicker, lines the quenched
cross sections.  The total cross sections are obtained by integrating
these curves.  We obtain the values of 36.0, 42.0 and 46.0 10$^{-16}$
fm$^2$ for the bare RPA calculations, and 7.0, 19.2 and 21.3
10$^{-16}$ fm$^2$ for the quenched calculations.  These results should
be compared with the experimental value of 14.0 $\pm$ 1.2 10$^{-16}$
fm$^2$ \cite{bod94}. The spreading of the various theoretical results
is much larger than the experimental uncertainty.

\vskip 0.5 cm 
In summary, we can conclude that the contribution of the excitation of
discrete states to the total neutrino-nucleus cross section is not
negligible, even for neutrino energies of a few hundred MeV. There are
large theoretical uncertainties in the description of the low-lying
discrete spectrum of medium-heavy nuclei, especially for unnatural
parity states. These uncertainties have large effects on the cross
sections of neutrinos of energies up to several tens of MeV, such as
the neutrinos coming from muon decay or supernovae neutrinos
\cite{bot05a}.

In order to obtain a satisfactory description of the low-lying
discrete spectrum, and, hopefully, to reduce the theoretical
uncertainties, it is necessary to use theories which goes beyond the
RPA framework by considering more complicated excitations, such as
$2p-2h$ degrees of freedom \cite{kre80,dro90,kam04}. Furthermore, in
the specific case of the $^{12}$C nucleus, there are indications that 
deformation effects are important \cite{krm05}.

\vskip 0.5 cm 
This work has been partially supported by the MURST through the PRIN:
{\sl Teoria della struttura dei nuclei e della materia nucleare}.


%
\end{document}